\documentclass[aps,showpacs,twocolumn,amsmath,amssymb]{revtex4}
\usepackage{amsmath,amssymb}
\usepackage{graphics,epsfig}
\usepackage{graphicx}
\usepackage{color}

\def\build#1_#2^#3{\mathrel{\mathop{\kern 0pt#1}\limits_{#2}^{#3}}}

 \newcommand{\vct}[1]{{\mbox {\boldmath $#1$}}}

\begin{document}

\title{Modeling the pressure Hessian and viscous Laplacian in Turbulence: comparisons with DNS and implications on velocity gradient dynamics}

\author{L. Chevillard$^{1,2}$, C. Meneveau$^1$, L. Biferale$^3$, F. Toschi$^4$}
\affiliation{$^1$Department of Mechanical Engineering and Center
for Environmental and Applied Fluid Mechanics, The Johns Hopkins
University, 3400 N. Charles Street, Baltimore, MD 21218,
USA\\$^2$Laboratoire de Physique de l'\'Ecole Normale Sup\'erieure
de Lyon, CNRS, Universit\'e de Lyon,  46 all\'ee d'Italie F-69007 Lyon,
France\\
$^3$Dipartimento di Fisica and INFN, Universit\`a di Roma ``Tor
Vergata'', Via della Ricerca Scientifica 1, 00133 Roma,
Italy.\\$^4$Istituto per le Applicazioni del Calcolo CNR, Viale
del Policlinico 137, 00161 Roma, Italy.}

\begin{abstract}
Modeling the velocity gradient tensor $\textbf{A}=\vct{\nabla}{\bf
u}$ along Lagrangian trajectories in turbulent flow requires
closures for the pressure Hessian and viscous Laplacian of
$\textbf{A}$. Based on an Eulerian-Lagrangian change of variables
and the so-called Recent Fluid Deformation closure, such models
were proposed recently (Chevillard \& Meneveau, Phys. Rev. Lett. \textbf{97}, 174501 (2006)).
The resulting stochastic model was shown to reproduce many
geometric and anomalous scaling properties of turbulence. In this
work, direct comparisons between  model predictions and  Direct
Numerical Simulation (DNS) data are presented. First, statistical
properties of $\textbf{A}$ are described using conditional
averages of strain skewness, enstrophy production, energy transfer
and vorticity alignments, conditioned upon invariants of the
velocity gradient.  These conditionally averaged quantities are
found to be described  accurately by the stochastic model.  More
detailed comparisons that focus directly on the terms being
modeled in the closures are also presented. Specifically,
conditional statistics associated with the  pressure Hessian and
the  viscous Laplacian are measured from the model and are
compared with DNS. Good agreement is found in strain-dominated
regions. However, some features of the pressure Hessian linked to
rotation dominated regions are not reproduced accurately by the
model. Geometric properties such as vorticity alignment  with
respect to principal axes of the pressure Hessian are mostly
predicted well. In particular, the model predicts that an
eigenvector of the rate-of-strain will be also an eigenvector of
the pressure Hessian, in accord with basic properties of the Euler
equations. The analysis identifies under what conditions the
Eulerian-Lagrangian change of variables with the Recent Fluid
Deformation closure works well, and in which flow regimes it
requires further improvements.
\end{abstract}


\maketitle

\section{Introduction}
\label{sec-intro} Fundamental understanding of universal features
of the small-scale structure of turbulence  has been a
long-standing challenge in turbulence research
\cite{K41,TenLum72,Kra74,MonYag75,Fri95,Pop00,TsinoBook}.  While
considerable phenomenological understanding has been accumulated
in recent decades,  the challenge of relating observed phenomena
and statistical properties to the dynamical equations
(Navier-Stokes) remains unmet.  The velocity gradient tensor
$A_{ij} = \partial u_i/\partial x_j$ (where $\textbf{u}$ denotes
the velocity vector) provides a rich characterization of the
topological and statistical properties of the fine-scale
structures in turbulence. Having a spectral peak at around the
Kolmogorov wavelength $k_\eta\sim\eta^{-1}$ ($\eta$ is the
Kolmogorov dissipative length scale),  $A_{ij}$ is a quantity
dominated  by motions in the viscous range. The antisymmetric part
of the tensor is the vorticity representing small-scale rotation
of fluid elements, while its symmetric part, the strain-rate
tensor, represents fluid deformation rate. The Lagrangian
evolution of this tensor can be described by an evolution equation
that is obtained from taking the gradient of the Navier-Stokes
equations. The resulting system is unclosed since it contains the
anisotropic part of the pressure Hessian and the viscous term. When
both of these are neglected, the system is closed (called
Restricted Euler - RE- dynamics)
\cite{LeoPhD,Vieille1,Vieille2,Cant1}. The RE equations already
predict several known geometric turbulence phenomena associated
with $A_{ij}$
\cite{Cant2,Ash87,Lun94,ooimar99,BosTao02,TaoKat02,Zef03,TsinoLag,GibHol07},
such as preferential alignments of vorticity with strain-rate
eigenvectors and preponderance of axisymmetric expansion and
positiveness of the intermediate eigenvalue of the strain-rate.
Neverteless, RE produces singularities in a relatively short,
finite time (see Ref. \cite{TsinoBook} for a review). 

Phenomenologically, other phenomena such as
small-scale intermittency may also  be probed, by studying the
probability distribution functions (PDF) of individual velocity
gradient elements. For instance, it is known that the PDFs of
longitudinal and transverse gradients, e.g. $A_{11}$ and $A_{12}$
in  particular directions, respectively, can be described by elongated stretched
exponential tails \cite{Fri95,CasGag90,KaiSre,CheCas06} or 
by superposition of stretched exponential \cite{bifprl}. Also the
moments of these gradients scale in non-trivial (anomalous) ways
with the Reynolds number \cite{Nel90,sreene}. 
Understanding such anomalous scaling behavior of turbulence is
another open challenge.  Therefore, the wealth of geometric,
dynamical and statistical turbulence phenomena that can be
described by the velocity gradient tensor, coupled with the fact
that a dynamical equation (even though unclosed) is available from
the gradient of the Navier-Stokes equations, makes $A_{ij}$ a
tensor variable of considerable interest for further study. The role of pressure in the intermittent nature of velocity gradients was also pointed out by Kraichnan in early works \cite{Kra90,Kra91}.

Based on prior works
\cite{Vieille2,Cant1,GirPop90,ChePum99,JeoGir03,LiMen05,LiMen06},
a stochastic dynamical model for the time evolution of  $A_{ij}$
has been proposed \cite{CheMen06,CheMen07CRAS}. The model includes
a closure for the pressure Hessian, i.e. $\partial ^2p/\partial
x_i\partial x_j$, and the viscous term, i.e. $\nu \nabla^2
\textbf{A}$ in terms of the local value of the velocity gradient
tensor. The approach, reviewed in detail in \S \ref{sec-theory},
consists in a change of variables from Eulerian positions to
Lagrangian labels before assuming isotropy in the associated
gradient tensors to be modeled. The Eulerian-Lagrangian
transformation involves a Jacobian matrix that is modeled using
the local value of the velocity gradient tensor by using the
``Recent Fluid Deformation'' closure. The model system is forced
using a Gaussian white-in-time random force. The resulting
stochastic model consists of eight independent coupled stochastic
differential equations (SDEs) that aim to describe the time evolution of
each of the tensor elements of $A_{ij}$, following a fluid
particle in a turbulent flow.

The results of Ref. \cite{CheMen06} show that the finite-time
divergence exhibited by the Restricted Euler system is regularized
with the inclusion of the proposed models for pressure Hessian and
viscous term. And, with the random forcing, stationary statistics of
the velocity gradient tensor are obtained, with realistic statistical
properties such as preferential alignment of vorticity and the
preferential state of axisymmetric expansion.  The shape of
probability distribution function of longitudinal and transverse
gradients is quite realistic and even some well-known properties of
anomalous scaling in turbulence are reproduced
\cite{CheMen06,CheMen07CRAS}.  A limitation of the model is that at
high Reynolds numbers the resulting distribution functions became
increasingly unrealistic. 
One approach to remedy this problem has been
explored \cite{BifChe07} by constructing a multi-scale version of the
model, i.e. a matrix shell model that describes the velocity gradient
tensors in various shells at different scales. 
The closure for the inter-scale interaction terms is based on the criterion
that the total kinetic energy must be preserved by
the modeled quadratic inter-scale interaction terms. 
The introduction of non-local (in scale) interactions leads to a structure of
the model that is more difficult to analyze theoretically, but it provides an interesting connection between the gradients' evolution at various scales and the energy cascade mechanism. 
While the matrix shell model successfully eliminates the problems at high Reynolds numbers,  it does not make an explicit connection with the physics of the pressure Hessian.  At this stage, then, it is of interest to further improve our understanding of the fundamental properties of the  closures proposed in \cite{CheMen06}, developed from the expression of pressure Hessian and viscous term as given from the Navier-Stokes equations, within the range of Reynolds numbers in which the model of \cite{CheMen06} works well.

In \S \ref{sec-firstcomp} various model predictions of statistical
and geometric properties of  $\textbf{A}$ beyond those already
studied in  \cite{CheMen06} are compared with Direct Numerical
Simulation (DNS) at a moderate Reynolds number ($\mathcal
R_\lambda=150$). The model is evaluated using statistical
measures already studied in \cite{ChePum99} in the context of
the ``tetrad model''. These measures include conditional averages of the
``dissipation" ($|\textbf S|^2 = S_{ij}S_{ij}$, where $\textbf{S}$
is the symmetric part of $\textbf{A}$) and of the  ``enstrophy''
(i.e. $|\vct{\Omega}|^2 = \Omega_{ij}\Omega_{ij}$, where
$\vct{\Omega}$ is the antisymmetric part of $\textbf{A}$). Then, a
similar analysis is performed with the enstrophy production and
the strain skewness.  The conditional averages are expressed in
terms of the  two principal invariants of $\textbf{A}$, namely
$R=-(1/3)\mbox{Tr}[\textbf{A}^3]$ and
$Q=-(1/2)\mbox{Tr}[\textbf{A}^2]$.  Different regions in the
``$(R,Q)$-plane'' have distinct physical interpretations
\cite{Cant1,Cant2,ChoPer90,ooimar99} and the behavior of the
computed conditional averages in these different regions thus
provide useful and statistically meaningful insights into the
performance of the model in dynamically very different regions of
the flow.

In order to quantify and understand the average local evolution of
the turbulence dynamics in the $(R,Q)$-plane,  the probability
current of the joint probability density $\mathcal P(Q,R)$ is also
studied in \S \ref{sec-flux}. These statistics  depend explicitly
on both pressure Hessian and viscous Laplacian, and thus the
effect of  the proposed closures for these terms may be compared
with the real effects obtained from the DNS.

In \S \ref{sec-alignments}  the preferential alignement of
vorticity with eigendirections of both pressure Hessian and the
symmetric part of the viscous Laplacian are studied in detail.
Connections with theoretical results pertaining to the Euler
equations are also made. Finally, in \S \ref{sec-conclusions} the
results are summarized and conclusions are presented.

\section{Theoretical background}
\label{sec-theory}
\subsection{Lagrangian description of the velocity gradient tensor}
A description of small-scale structure of turbulence based on the
velocity gradient tensor $A_{ij} = \partial u_i/\partial x_j$
begins by taking the gradient of the Navier-Stokes equation. One
then obtains the system:
\begin{equation}\label{eq:NS}
\frac{dA_{ij}}{dt} = -A_{ik}A_{kj}-\frac{\partial^2p}{\partial
x_i\partial x_j}+\nu\frac{\partial^2A_{ij}}{\partial x_k\partial
x_k}\mbox{ ,}
\end{equation}
where $d/dt$ stands for the Lagrangian material derivative (i.e.
$d/dt\equiv \partial/\partial t+u_k\partial/\partial x_k$), $p$
the pressure divided by the density of the fluid and $\nu$ the
kinematic viscosity.  Because of incompressibility, $\textbf{A}$
must remain trace-free, i.e. $A_{ii}=0$. Equation \ref{eq:NS} is
not closed in terms of $\textbf{A}$ at the position $\textbf{x}$
and time $t$. This can be easily seen noting that the pressure
field is the solution of the Poisson equation
$\mbox{Tr}[{\partial^2p}/{\partial x_i\partial x_j}] = \nabla^2
p=-A_{lk}A_{kl}$ which shows that pressure is highly non-local.
And the viscous term requires the Laplacian of $\textbf{A}$ which
is not known simply in terms of $\textbf{A}$.

As already mentioned in \S \ref{sec-intro}, neglecting pressure
Hessian anisotropy and viscous effects leads to finite time
singularities because of the strong and unopposed effects of the
self-streching term $-\textbf{A}^2$. One can find in the
literature several efforts at regularizing this finite time
divergence, while keeping the exact self-streching term. Firstly,
Girimaji and Pope \cite{GirPop90} succeded to do so by
constructing a stochastic model with an imposed constraint. This
constraint is imposed by modifying the non-linear term so that the
pseudo-dissipation $\varphi=A_{ij}A_{ij}$ \cite{PopChe90}  is
lognormal with a prescribed mean and variance.  Intermittency
trends are put in  explicitly, by prescribing a known variance of
$\log(\varphi)$ as function of the Reynolds number.

More recently, two groups proposed the idea that the local
geometry of the accumulated  fluid deformation, i.e. formally the
Cauchy-Green tensor, may represent the missing information which
allows to regularize the RE divergence. Accumulated  fluid
deformation thus has been used to model the pressure Hessian in
the so-called  ``tetrad model'' of Chertkov, Pumir and Shraiman
\cite{ChePum99,NasPum05,NasPum07}. A similar idea from Jeong and
Girimaji \cite{JeoGir03} has been used to model the viscous part
of Eq. (\ref{eq:NS}), explicitly using the Cauchy-Green tensor.
Whether or not the finite time divergence is regularized, the
direct use of the Cauchy-Green tensor is limited by the fact that
this tensor is fundamentaly non-stationary, i.e. as time evolves
it maintains memory of the initial condition. Hence the resultant
models for pressure Hessian and viscous term are intrinsically
non-stationary and depend on the initial condition chosen to
initialize the material deformation tracking.  In the following
sections, we discuss these issues in more detail and also review
the simplified model of \cite{CheMen06} that avoids these problems
of non-stationarity.

We also point out an alternative approach \cite{cantwell01} 
that renormalizes the time variable in RE dynamics so as to 
relegate the finite time singularities to infinite time.   

\subsection{Lagrangian mapping and Cauchy-Green Tensor}

Following Refs. \cite{MonYag75,Con01,MajBer02}, one may define a
mapping $\mathcal{T}_{t_0,t}$ between Eulerian and Lagrangian
coordinates:
\begin{equation}\label{eq:Mapping}
\mathcal{T}_{t_0,t}:\textbf{X}\in \mathbb R^3\mapsto
\textbf{x}\in \mathbb R^3\mbox{ ,}
\end{equation}
where $\textbf{x}(\textbf{X},t)$ denotes the position at a time
$t$ of a fluid particle which was at the position
$\textbf{x}(\textbf{X},t_0)=\textbf{X}$ at the initial time $t_0$.
Given the initial position of a fluid particle, this mapping (Eq.
(\ref{eq:Mapping})) is fully defined at any time by
\begin{equation}\label{eq:DetODE}
\frac{d\textbf{x}}{dt}=\textbf{u}(\textbf{x},t)\mbox{ .}
\end{equation}
A quantity of much interest in continuous mechanics  is the
deformation gradient tensor \textbf{D}, defined as $D_{ij} =
\partial x_i/\partial X_j$, which relates the variation of the
position of a particle when one slightly changes the initial
position. Differentiating Eq. (\ref{eq:DetODE}) with respect to
$X_j$, one gets the time evolution of \textbf{D}, i.e.
\begin{equation}\label{eq:DynEvolD}
\frac{d\textbf{D}}{dt} = \textbf{A}(t)\textbf{D}(t)\mbox{ ,}
\end{equation}
and one can show \cite{MonYag75,MajBer02} that the  Jacobian of
the mapping $\mathcal{T}_{t_0,t}$, i.e.
$\mbox{det}(\textbf{D}(t))$, is equal to unity at any time by
virtue of incompressibility, stating that this mapping is always
invertible. Eq. (\ref{eq:DynEvolD}) can be exactly solved using
the product integral \cite{DolFri79} or alternatively, the
time-ordered exponential \cite{ItzZub80,FalGaw01}
\begin{equation}\label{eq:TimeOrdered}
\textbf{D} = \prod_{t_0}^t e^{ds\textbf{A}(s)} =  \mathcal
T^+\exp\left[ \int_{t_0}^tds\textbf{A}(s)\right]\mbox{ .}
\end{equation}
The Cauchy-Green tensor, $\textbf{C}(t)$, is defined as the
symmetric tensor $\textbf{C} = \textbf{D}\textbf{D}^\top$ and its
eigenvalue and eigenvector system  describes the rotation and
deformation of initially isotropic-shaped fluid volumes into
various shapes as time goes on. The transport equation of the
Cauchy-Green tensor can be obtained in straightforward fashion
\cite{TruNol92}  from Eq. (\ref{eq:DynEvolD}):
\begin{equation}\label{eq:DynEvolC}
\frac{d\textbf{C}}{dt} = \textbf{A}(t)\textbf{C}(t) +
\textbf{C}(t)\textbf{A}^\top(t)\mbox{ .}
\end{equation}
Based on the properties of  $\textbf{C}$,  studies of isotropic
and homogeneous turbulence in both numerical \cite{GirPop90Num}
and laboratory \cite{TsinoLag,TsinoBook} flows have shown that
cigar (one large and two small eigenvalues of $\textbf{C}$) and
pancake (two large and one small eigenvalue) shapes are the most
common shapes of fluid deformation.

\subsection{Fluid deformation, and pressure Hessian models}

Let us first remark that the pressure Hessian is not among the
most studied objects in the turbulence literature (although, see
Ref. \cite{NomPos98}). One reason perhaps is that it cannot be
described naturally from a standard transport equation along a
Lagrangian trajectory. Instead, the pressure Hessian is related to
the spatial distribution of the velocity gradient using singular
integral operators \cite{Ohk93,OhkKis95,MajBer02}:
\begin{equation}\label{eq:SOPH}
\frac{\partial^2 p}{\partial x_i\partial x_j} =
-\mbox{Tr}(\textbf{A}^2)\frac{\delta_{ij}}{3}-\mbox{P.V.}\int
k_{ij}(\textbf{x}-\textbf{y})
\mbox{Tr}(\textbf{A}^2)(\textbf{y})d\textbf{y}
\end{equation}
where the integral is understood as a Cauchy principal value
(P.V.) and $k_{ij}$ is the Hessian of the Laplacian's Greens
function, namely
\begin{equation}
k_{ij}(\textbf{x}) = \frac{\partial ^2}{\partial x_i\partial x_j}
\frac{1}{4\pi|\textbf{x}|} =
\frac{|\textbf{x}|^2\delta_{ij}-3x_ix_j}{4\pi|\textbf{x}|^5}
\mbox{ .}
\end{equation}
One can see from Eq. (\ref{eq:SOPH})  that only the isotropic part
of the pressure Hessian is purely local (the first term of the RHS
of Eq. (\ref{eq:SOPH})). All the nonlocal effects of pressure
Hessian  enter through the anisotropic part (or deviatoric part
corresponding to the second term of the RHS of Eq. (\ref{eq:SOPH})).
Hence, in this view, the RE approximation can be understood as the
neglect of all the nonlocal effects implied by the
incompressiblity condition: the corresponding Lagrangian particle
evolves with the flow completely independent from its neighbors.
As far as we know, the tetrad model \cite{ChePum99} is the first
model to have been proposed for the anisotropic (i.e. nonlocal)
part of the pressure Hessian. While the authors introduced the
model using the language of multipoint dispersion of particles
that define an evolving tetrad shape, a simple interpretation of
the model can also be given in terms of the deformation and
Cauchy-Green tensors.

To begin, one can re-express various Eulerian quantities  such as
the pressure Hessian and the viscous term in terms of Lagrangian
coordinates, i.e. in terms of the fluid particle's position at
some initial time $t_0$, $\textbf{X}$. For the  Hessian tensor of
the pressure at the current point and time $(\textbf{x},t)$ one
may write
\begin{equation}\label{eq:ChangVar}
\frac{\partial ^2p(\textbf{x},t)}{\partial x_i\partial x_j} = \frac{\partial
X_p}{\partial x_i}\frac{\partial X_q}{\partial x_j}\frac{\partial
^2p(\textbf{x},t)}{\partial X_p\partial X_q} + \frac{\partial^2 X_q}{\partial
x_i\partial x_j}\frac{\partial p(\textbf{x},t)}{\partial X_q}\mbox{ .}
\end{equation}

The second term entering in the RHS of Eq.
(\ref{eq:ChangVar}) requires the knowledge of
the spatial distribution of the (inverse) deformation gradient,
through its spatial derivative.  As will be seen later, the adopted
approach neglects short-time variations in the velocity gradient
and in the context of the proposed Lagrangian model, 
it is natural to neglect spatial fluctuations of the deformation gradient,
i.e. ${\partial^2 X_q}/{\partial x_i\partial x_j} \sim 0$. 
Next, we discuss the remaining term of the RHS of Eq.
(\ref{eq:ChangVar}). The fourth-order tensor $\partial_i X_p
\partial_j X_q$ can be solved   along its trajectory using the
dynamical evolution for the deformation tensors (Eq.
(\ref{eq:DynEvolD})). For the remaining factor, the Lagrangian
pressure Hessian ${\partial ^2p }/{\partial X_p\partial X_q}$, we
choose the simplest assumption, namely the isotropic assumption:
\begin{equation}\label{eq:IsoPressHess}
\frac{\partial ^2p}{\partial X_p\partial X_q}
 =  \frac{1}{3}\frac{\partial ^2p}{\partial X_m\partial X_m}
\delta_{pq}\mbox{ .}
\end{equation}
Physically, this assumption states that as time progresses, one
looses memory about the relative orientations of the initial
locations  $\textbf{X}$ as far as the present value of pressure is
concerned. The contraction between  $\delta_{pq}$ and $\partial_i
X_p \partial_j X_q$ then connects the model to the Cauchy-Green
tensor introduced in the preceding section.

So far the pressure Hessian can then be rewritten, using Eq.
(\ref{eq:IsoPressHess}), according to
\begin{equation}\label{eq:ChangVar2}
\frac{\partial ^2 p}{\partial x_i \partial x_j } \approx
\frac{\partial X_m}{\partial x_i}\frac{\partial X_n}{\partial
x_j}\frac{\partial ^2 p(\textbf{x},t)}{\partial X_m \partial X_n
}~= C^{-1}_{ij}\frac{1}{3}\frac{\partial
^2p(\textbf{x},t)}{\partial X_k\partial X_k}\mbox{ .}
\end{equation}

To determine ${\partial ^2p}/{\partial X_k\partial X_k}$, we
follow Ref. \cite{ChePum99} and use the Poisson equation
$\nabla^2p=-A_{nm}A_{mn}=(1/3)C^{-1}_{qq}{\partial ^2p}/{\partial
X_k\partial X_k}$. Replacing back into Eq. (\ref{eq:ChangVar2}),
leads to \cite{ChePum99,CheMen06}:
\begin{equation}\label{eq:ModHessTetrad}
\frac{\partial ^2p(t)}{\partial x_i\partial x_j} =
-\frac{\mbox{Tr}(\textbf{A}^2)}{\mbox{Tr}
(\textbf{C}^{-1})}C_{ij}^{-1}= \frac{2
Q}{\mbox{Tr}(\textbf{C}^{-1})}~C_{ij}^{-1}\mbox{ .}
\end{equation}

\subsection{Fluid deformation and modeling the viscous Laplacian}

In a similar fashion, following Ref. \cite{JeoGir03}, this
procedure can be applied to the viscous Laplacian entering in the
gradient of the Navier-Stokes equation (Eq. (\ref{eq:NS})), i.e.
\begin{equation}\label{eq:LagHessA}
\nu\frac{\partial ^2\textbf{A}}{\partial x_k\partial x_k} \approx
\frac{\partial X_p}{\partial x_k}\frac{\partial X_q}{\partial
x_k}\left(\nu\frac{\partial ^2\textbf{A}}{\partial X_p\partial
X_q} \right)\mbox{ .}
\end{equation}
The resulting Lagrangian Hessian of \textbf{A} entering in Eq.
(\ref{eq:LagHessA}) will be considered as (i) isotropic, i.e.
$\partial ^2\textbf{A}/(\partial X_p\partial X_q) = \partial
^2\textbf{A}/(\partial X_m\partial X_m)\delta_{pq}/3$ and (ii),
its trace will be modeled by a friction term, i.e. $\partial
^2\textbf{A}/(\partial X_m\partial X_m) = -1/\ell^2\textbf{A}$.
The characteristic length scale $\ell$ reflects the typical length
in the Lagrangian frame over which $\textbf{A}$ is correlated. To
estimate this length-scale, we note that the typical decorrelation
time of $\textbf{A}$ along its Lagrangian trajectory is known to
be on the order of  $\tau_K=(\nu/\epsilon)^{1/2}$, the Kolmogorov
time-scale (where $\epsilon$ is the dissipation rate)
\cite{TenLum72}. During that time, a fluid particle is advected
by the turbulence over a distance of the order of 
$\ell = u^\prime \tau_K =\lambda$, where
$u^\prime$ is the root mean square velocity (chosen as advective
velocity scale) and $\lambda$ the Taylor microscale. Finally, recognizing
that $\nu/\lambda^2 = T^{-1}$, where $T$ is the integral time
scale, one then obtains the following model for the viscous term:
\begin{equation}\label{eq:ModViscA}
\nu\frac{\partial ^2\textbf{A}}{\partial x_k\partial x_k} \approx
- \frac{1}{T} \frac{\mbox{Tr}(\textbf{C}^{-1})}{3}\textbf{A}\mbox{
.}
\end{equation}
This model is similar to the one obtained by Jeong and Girimaji
\cite{JeoGir03} but using a different, more physically motivated
time scale.

\subsection{Stochastic model based on the Recent Fluid Deformation (RFD) closure}

The various terms entering in the rhs. of Eqs.
(\ref{eq:ModHessTetrad}) and (\ref{eq:ModViscA}) include the
tensor $\textbf{C}$. If this tensor is obtained from its transport
equation (Eq. (\ref{eq:DynEvolC})) subject to the natural initial
condition $C_{ij}(t_0) = \delta_{ij}$, then the closures for
pressure Hessian and viscous term depend strongly on the initial
time $t_0$, or equivalently, on the initial position \textbf{X}.
Due to the dispersive nature of turbulent flow, $\textbf{C}$
continues to evolve with exponentially growing and decreasing
eigenvalues. Instead of solving for $\textbf{C}$ from its
transport equation and having to deal with the problems associated
with non-stationarity, in \cite{CheMen06} a simple closure was
proposed. It consists of a sort of `Markovianization' of the
dynamics of $\textbf{C}$ in that it is assumed that $\textbf{C}$
evolves in a frozen velocity gradient tensor field during a
characteristic (short) time $\tau$. The value of  $\textbf{A}$
during that time is taken as the most recent value (i.e. the
current, local, value). And  the time-scale chosen is the typical
de-correlation time-scale of  $\textbf{A}$ during its Lagrangian
evolution, which is known to be of the order of the Kolmogorov
time-scale, $\tau_K$.  Thus, the initial time is taken to be at
$t_0=t-\tau_K$, which allows to write in a simple way the time
ordered exponential entering in Eq. (\ref{eq:TimeOrdered}). We
thus replace the true  Cauchy-Green tensor  by a new tensor,
called the ``recent Cauchy-Green tensor' $\textbf{C}_{\tau_K} $
that can be expressed in terms of simple matrix exponentials:
\begin{equation}\label{eq:MyCG}
\textbf{C}_{\tau_K} = e^{\tau_K \textbf{A}}e^{\tau_K
\textbf{A}^\top}\mbox{ .}
\end{equation}
This leads to an explicit $\textbf{A}$-dependent model for the
full pressure Hessian:
\begin{equation}\label{eq:modHess}
\frac{\partial^2 p}{\partial x_i\partial x_j} =
-\frac{\mbox{Tr}(\textbf{A}^2)}{\mbox{Tr}(\textbf{C}_{\tau_K}^{-1})}
\left(\textbf{C}_{{\tau_K}}^{-1}\right)_{ij}
\end{equation}
and for the viscous Laplacian
\begin{equation}\label{eq:RFDModViscA}
\nu\frac{\partial ^2\textbf{A}}{\partial x_k\partial x_k} = -
\frac{1}{T}
\frac{\mbox{Tr}(\textbf{C}_{\tau_K}^{-1})}{3}\textbf{A}\mbox{ .}
\end{equation}

Inserting Eqs. (\ref{eq:modHess}) and (\ref{eq:ModViscA}) into Eq.
(\ref{eq:NS}) and writing the equation in the It\^{o}'s language of
stochastic differential equations \cite{KloPla92} the full model
for the time evolution of the velocity gradient reads
\begin{equation}\label{eq:Determourmodel}
d\textbf{A} =  \left(-\textbf{A}^2+
\frac{\mbox{Tr}(\textbf{A}^2)}{\mbox{Tr}(\textbf{C}_{{\tau_K}}^{-1})}
\textbf{C}_{{\tau_K}}^{-1}
-\frac{\mbox{Tr}(\textbf{C}_{{\tau_K}}^{-1})}{3T}
\textbf{A}\right)dt+d\textbf{W}\mbox{ .}
\end{equation}
The stochastic time evolution of the velocity gradient tensor $\textbf{A}$ (Eq. \ref{eq:Determourmodel}), as proposed in Refs. \cite{CheMen06,CheMen07CRAS}, relates the joint deterministic action of the self-streching term $-\textbf{A}^2$, the pressure Hessian (Eq. \ref{eq:modHess}) and the viscous term (Eq. \ref{eq:RFDModViscA}).
Moreover, the system is forced with a stochastic Gaussian noise. The deterministic part provides two time scales: a small time scale $\tau_K$ and a large one $T$. The latter arises in modeling the viscous diffusion
term when combining the viscosity with the Taylor-microscale, which in turn is related 
to the large-scale velocity rms.  Hence, the deterministic part gives the dependence on the Reynolds number $\mathcal R_e$ to the model through the ratio $(T/\tau_K)^2\sim \mathcal R_e$, according to classical Kolmogorov dimensional arguments \cite{Fri95}. Dependence on the Reynolds number of higher order moments of velocity derivatives (i.e. anomalous scalings and the intermittency phenomenon) have been studied and quantified in Ref. \cite{CheMen07CRAS}. The purpose of this article is to focus on a single Reynolds number and to compare it with a DNS flow (see next paragraph).

The term $\textbf{W}$ is a tensorial delta-correlated noise term
that has been added in order to represent possible forcing
effects, e.g. from neighboring eddies
\cite{CheMen06,CheMen07CRAS}. In Appendix \ref{ann:DefBruit}, we
describe this noise extensively, and propose a way to simulate it.

\subsection{DNS data and comparisons with the model}

In the following, we will make extensive use of a standard direct
numerical simulation (DNS) of the Navier-Stokes equation for a
Taylor based Reynolds number of order $\mathcal R_\lambda=150$.
Pseudo-spectral simulations are performed, of an isotropic
turbulent flow in a $[0,2\pi]^3$ box using $256^3$
nodes. Fourier modes in shells with $|{\bf k}| < 2$ are forced by
a term added to the Navier-Stokes equations, which provides
constant energy injection rate $\epsilon_f = 0.1$. The viscosity
of the fluid is $\nu=0.00113$. The time step $\Delta t$ is chosen
adaptively to ensure the Courant number $\Delta t u_{\rm
max}/\Delta x\leqslant 0.15$, where $u_{\rm max}$ is the maximum
velocity and $\Delta x$ is the grid size. In order to make
comparisons between DNS data and the model, one has to specify a
value for the parameter of the model $\tau_K$. At $\mathcal
R_\lambda=150$, it has been estimated by Yeung \textit{et al.}
\cite{YeuPop06JoT} that the ratio of the Kolmogorov scale and the
integral (i.e. velocity correlation time scale) time scale is
$\tau_K/T \approx 0.1$. Thus, in the following, DNS data will be
compared to the model run with ${\tau_K} =0.1T$. Without loss of
generality, the integral time scale $T$ will be set to unity. It
corresponds to set time as units of $T$. The model as written out
as in Eq. (\ref{eq:Determourmodel}) is solved numerically, with the
parameter $\tau_K=0.1$, using a second order predictor-corrector
method (see Ref. \cite{KloPla92}) with a time step of $\Delta
t=10^{-3}$. One obtains time-series of all of the components of
the tensor $\textbf{A}$ that display temporally stationary
statistics. In this article, we have worked with a time-series of
length $\sim 10^6$ in units of the integral time scale $T$. These
can then be directly compared to DNS results. Furthermore the
model provides statistically stationary time-series for both
pressure Hessian and viscous Laplacian that can be also directly
compared to DNS data.

\section{Conditional statistics of the velocity gradient tensor}
\label{sec-firstcomp}

\begin{figure}[t]
\center{\epsfig{file=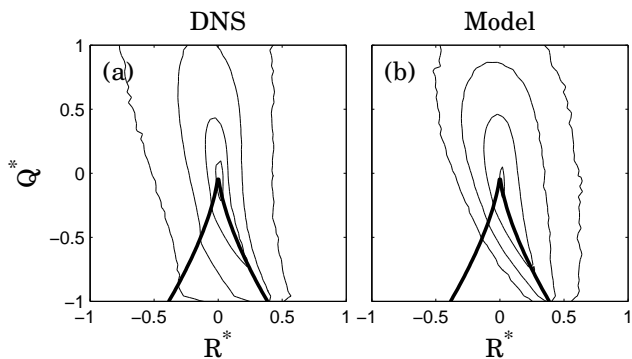,width=8.5cm}} \caption{Joint
PDF $\mathcal P(Q^*,R^*)$ of $R^* = R/\langle
S_{ij}S_{ij}\rangle^{3/2}$ and $Q^*= Q/\langle
S_{ij}S_{ij}\rangle$ calculated from DNS (a) and the present model
(b). Contour lines are the same in the two cases, logarithmically
spaced by a factor of 10, and start at 10 near the origin. The
thick line represents the zero-discriminant (or Vieillefosse)
line: $\frac{27}{4}R^2+Q^3=0$. } \label{fig:RQ}
\end{figure}

We are interested here in studying various properties of the
velocity gradient tensor conditioned upon the value of its two
invariants  $R$ and $Q$ defined earlier.  The joint probability
density of $(R,Q)$ (the  $RQ$-plane)  has been studied in the past
\cite{ChoPer90,Vieille2,Cant1,Cant2,ooimar99} and can be used to
characterize the frequency of occurrence of  the various local
topologies of the flow.

For instance, in a simple way, the second invariant
\begin{equation}\label{eq:DefQ}
Q= -\frac{1}{2}\mbox{Tr}({\textbf{A}}^2) = \frac{1}{4}
|\mathbf{\omega}|^2-\frac{1}{2} \mbox{Tr}(\textbf{S}^2)
\end{equation}
can be understood as the competition between enstrophy
($\mathbf{\omega}$ denotes vorticity) and dissipation (per unit
viscosity). Then, positive $Q$ represents rotation-dominated
regions and negative $Q$ dissipation-dominated regions.
Analogously, the third invariant
\begin{equation}\label{eq:DefR}
R= -\frac{1}{3}\mbox{Tr}(\textbf{A}^3) = -\frac{1}{4}
\omega_iS_{ij}\omega_j-\frac{1}{3} \mbox{Tr}(\textbf{S}^3)
\end{equation}
represents competition between \textit{enstrophy production},
entering in the enstrophy evolution \cite{TsinoBook,RosMen06}, i.e.
\begin{equation}\label{eq:EnstrEvol}
\frac{1}{2}\frac{d|\mathbf{\omega}|^2}{dt} =
\omega_iS_{ij}\omega_j +\nu\omega_i \nabla^2\omega_i\mbox{ ,}
\end{equation}
and the dissipation \textit{production} (or Strain skewness
\cite{ChePum99}) entering in the dissipation evolution , i.e.
\begin{equation}\label{eq:DissEvol}
\frac{d\mbox{Tr}(\textbf{S}^2)}{dt} = -2
\mbox{Tr}(\textbf{S}^3)-\frac{1}{2}\omega_iS_{ij}\omega_j -
2S_{ij}\frac{\partial ^2 p}{\partial x_i\partial x_j} +\nu S_{ij}
\nabla^2 S_{ij}\mbox{ .}
\end{equation}

Let us remark that one may interpret the RQ plane in a different
way, based on the eigenvalues of $\textbf{A}$ (two of them can be the
complex conjugate) and the zero-discriminant line (i.e. the
``Vieillefosse" line, namely $\frac{27}{4}R^2+Q^3=0$). See for
instance \cite{ChoPer90,Cant1,Cant2}.

\subsection{The joint PDF in the RQ-plane}
\begin{figure}[t]
\center{\epsfig{file=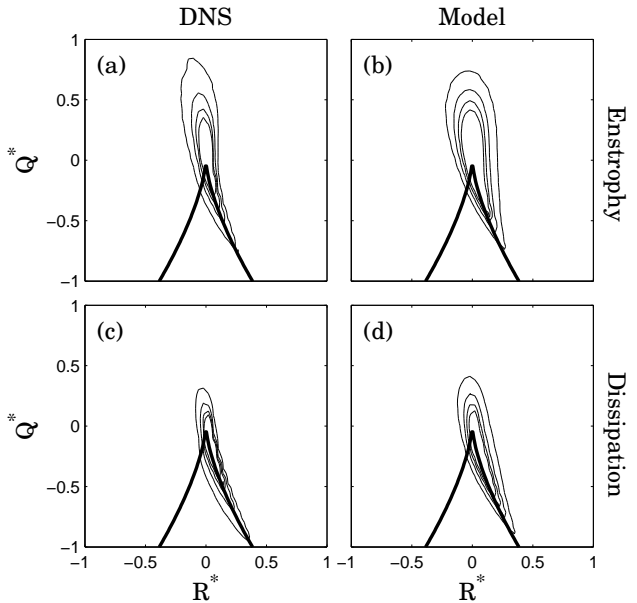,width=8.5cm}}
\caption{Isocontours of the conditional enstrophy
$\langle\omega_i\omega_i|Q^*,R^*\rangle \mathcal P(Q^*,R^*)$ where
$\vct{\omega}$ is the vorticity, and conditional dissipation
$\langle S_{ij}S_{ij}|Q^*,R^*\rangle \mathcal P(Q^*,R^*)$.
Following Ref. \cite{ChePum99}, both quantities are renormalized
by $\langle 2\Omega_{ij} \Omega_{ij} \rangle$, where
$\vct{\Omega}$ is the rate of rotation tensor. Level of contour
lines are  0.15, 0.3, 0.45, 1, 2, 3, and 4.} \label{fig:RQEnsDiss}
\end{figure}

We show in Fig. \ref{fig:RQ} the joint PDF of $R$ and $Q$, or
equivalently the joint PDF $\mathcal P(Q^*,R^*)$ of the
non-dimensionalized invariants $R^* = R/\langle
S_{ij}S_{ij}\rangle^{3/2}$ and $Q^*= Q/\langle
S_{ij}S_{ij}\rangle$, for both DNS and the model (Eq.
(\ref{eq:Determourmodel})). In order to compute various conditional averages, 
a range $[-1;1]$ of values of the two non-dimensionalized invariants $R^*$ and $Q^*$
of \textbf{A} is discretized in 25 equally spaced bins.  
For the DNS, one observes the 
predominance of the Enstrophy-Enstrophy production quadrant
($R^*<0$ and $Q^*>0$) and the Dissipation-Dissipation production
quadrant ($R^*>0$ and $Q^*<0$). The predominance of these two
quadrants has been observed before in the literature \cite{Cant2}.
The model reproduces these basic trends fairly accurately, with
the characteristic ``tear-shape" elongation along the
``Vieillefosse tail'' in the Dissipation-Dissipation production
quadrant. But one also observes that the model overestimates the
total probability in the Enstrophy-Dissipation production region
(i.e. $R^*>0$ and $Q^*>0$) and underestimates the
Dissipation-Enstrophy production region (i.e. $R^*<0$ and
$Q^*<0$). It will be shown later that  this is caused by
limitations in how the pressure Hessian is closed and modeled.

Next, we check whether or not dissipation (resp. enstrophy) is
dominantly associated with the $R^*>0$ and $Q^*<0$ (resp. $R^*<0$
and $Q^*>0$) quadrants. Following the approach already used in
\cite{ChePum99} we present in Fig. \ref{fig:RQEnsDiss} the
conditional averages of dissipation, i.e.
$\langle\mbox{Tr}(\textbf{S}^2)|Q^*,R^*\rangle \mathcal
P(Q^*,R^*)$ and enstrophy $\langle\omega_i\omega_i|Q^*,R^*\rangle
\mathcal P(Q^*,R^*)$ on $R^*$ and $Q^*$. Averages are weighted by
the joint density $\mathcal P(Q^*,R^*)$ to ensure that the sum
over all possible values of $R^*$ and $Q^*$ gives the averages of
respectively dissipation and enstrophy.  
We clearly see that the quadrants $R^*>0$; $Q^*<0$
(or $R^*<0$; $Q^*>0$) are dominated by dissipation (or enstrophy,
respectively). The model reproduces these conditional averages
quite accurately.

\subsection{Enstrophy production, Strain Skewness and Energy transfer}

\begin{figure}[t]
\center{\epsfig{file=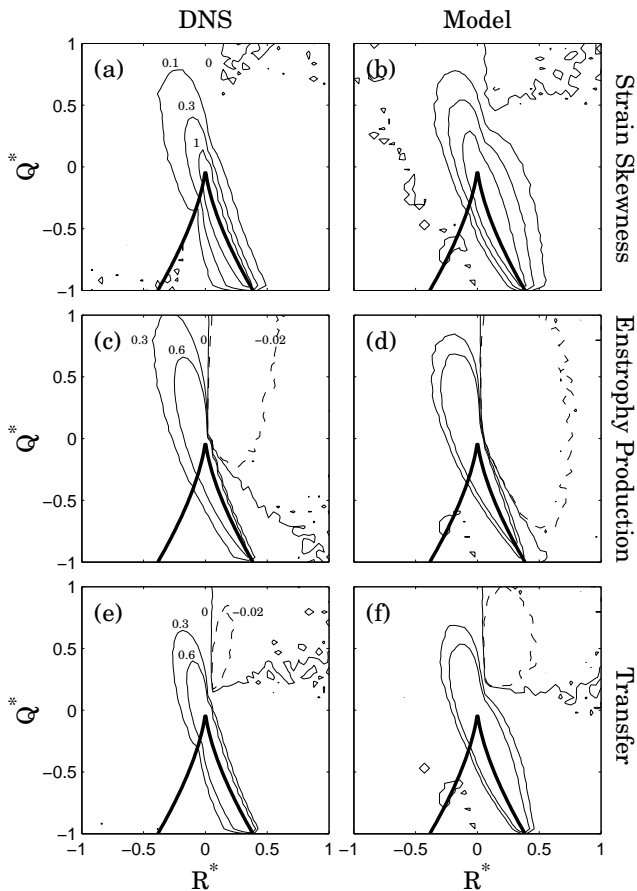,width=8.5cm}}
\caption{Isocontours of the conditional Strain Skewness
$-\langle\mbox{Tr}(\mbox{\textbf{S}}^3)|Q^*,R^*\rangle \mathcal
P(Q^*,R^*) $, enstrophy production $\langle \omega_i
S_{ij}\omega_j|Q^*,R^*\rangle \mathcal P(Q^*,R^*) $ and energy
transfer
$-\langle\mbox{Tr}(\mbox{\textbf{A}}^2\mbox{\textbf{A}}^T)|Q,R\rangle
\mathcal P(Q^*,R^*) $. Following Ref. \cite{ChePum99}, various
quantities are normalized by the average transfer
$|\langle\mbox{Tr}(\mbox{\textbf{A}}^2\mbox{\textbf{A}}^T)\rangle|$.
} \label{fig:RQTrans}
\end{figure}

A similar study is performed with the various quantities entering
in the definition of the third invariant $R$ (Eq. (\ref{eq:DefR})),
namely the enstrophy production and the strain skewness
\cite{ChePum99}. In Fig. \ref{fig:RQTrans} (a) to (d)  these
various quantities are shown, together with the predictions from
the stochastic model. In all cases, it is apparent that model
predictions are quite accurate. We see also that enstrophy
production is clearly dominant in the $R^*<0$ and $Q^*>0$
quadrant. Let us mention that in the enstrophy-dissipation
production dominated region ($R^*>0$ and $Q^*>0$), entrophy
production becomes weakly negative, stating that in this region,
enstrophy decreases with time (see Eq. (\ref{eq:EnstrEvol})). 
Also, in Fig. \ref{fig:RQTrans} (a-b), we
see that strain skewness is dominating in the bottom-right
quadrant, but remains very important in the top-left quadrant.
This is mainly linked to the fact that  the
evolution of dissipation not only depends on the strain skewness (or
dissipation production), but also on enstrophy production, and a term
linked to the pressure Hessian (see Eq. (\ref{eq:DissEvol})). 

A related quantity of interest is 
\begin{equation}\label{eq:EnerTrans}
-\mbox{Tr}(\textbf{A}^2\textbf{A}^\top) =-\mbox{Tr}(\textbf{S}^3)
-\frac{1}{4} \omega_iS_{ij}\omega_j
\end{equation}
which describes the time evolution of the pseudodissipation
$d\mbox{Tr}(\textbf{A}\textbf{A}^\top)/dt =
-\mbox{Tr}(\textbf{A}^2\textbf{A}^\top)$ in the RE approximation.
This quantity is sometimes called ``energy transfer''
\cite{Eyink95,BorOrz98,ChePum99}  when $\textbf{A}$ is defined by
filtering in the inertial range in the context of large eddy
simulations (see \cite{MenKatRev}). While here $\textbf{A}$ is not
filtered and therefore no such direct physical interpretation is available,
this quantity is still presented as additional documentation of
the properties of  $\textbf{A}$.  Results are displayed in Fig.
\ref{fig:RQTrans}(e-f). Once again, the model reproduces well the
trends observed in DNS, including negative regions in the
top-right quadrant.

\subsection{Geometric alignments of vorticity with strain-rate eigenvectors}

An important universal feature of fully developed turbulent flows
is the preferential alignment of vorticity along the
eigendirection of the intermediate eigenvalue of the strain-rate
tensor $\textbf{S}$ (see \cite{TsinoBook} and references therein).
To study the alignment
properties of vorticity conditioned on various values of $R$ and
$Q$ the  $(R,Q)$ plane is divided
into four regions related to the
eigenvalue structure of $\textbf{A}$. Instead of the $Q=0$ line to
separate high and low rotation regions as was done in the
qualitative discussions of the previous sections, we now use the
quantitatively more precise classification, in which the $(R,Q)$
plane is divided into high and low rotation regions by the
zero-discriminant line, i.e. $Q=-(27R^2/4)^{1/3}$.

\begin{figure}[h!]
\center{\epsfig{file=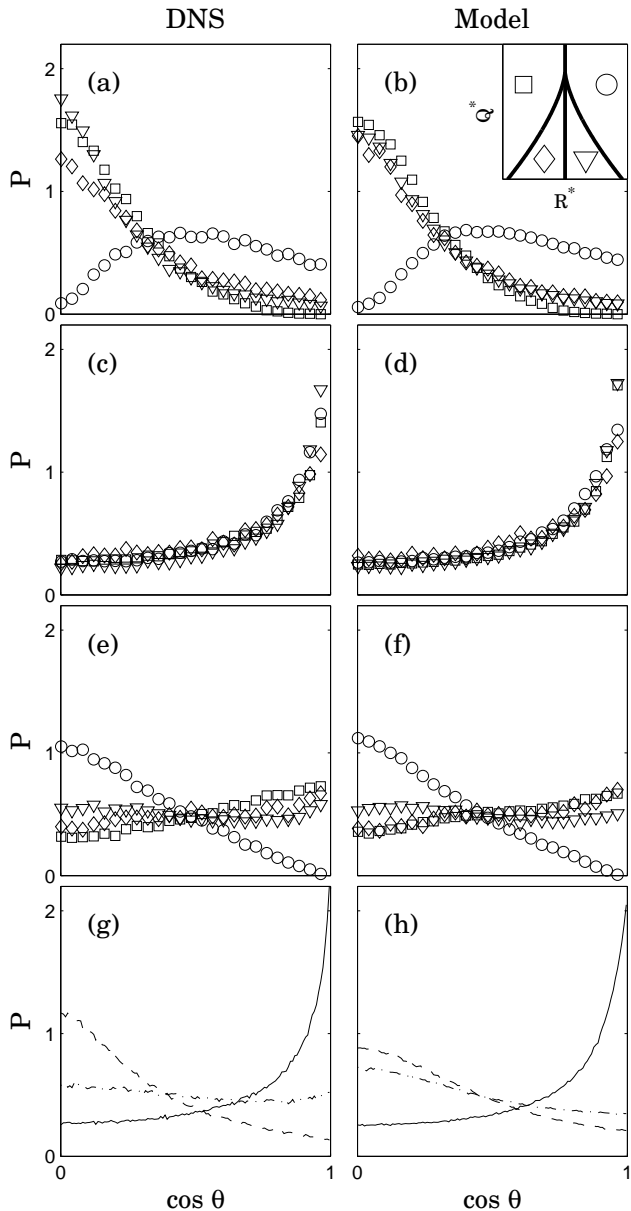,width=8.5cm}} \caption{PDF of
the cosine of the angle $\theta$ between vorticity and the
different eigendirections of $\textbf{S}$: (a) and (b) for the
most contractive eigendirection (negative eigenvalue), (c) and (d)
for the intermediate eigendirection, and (e) and (f) for the most
positive (extensive) eigen-direction. As it is schematically displayed in the inset of Fig. \ref{fig:RQGeoMin}(b), different symbols are
obtained from the four different regions of the $(R,Q)$ plane delimited by the $R>0$ and the \textit{Vieillefosse} (or zero-discriminant) lines given by $Q=-(\frac{27}{4} R^2)^{1/3}$: $\circ$
($R>0$ and $Q>-(\frac{27}{4} R^2)^{1/3}$), $\triangledown$ ($R>0$
and $Q<-(\frac{27}{4} R^2)^{1/3}$), $\square$ ($R<0$ and
$Q>-(\frac{27}{4} R^2)^{1/3}$), $\lozenge$ ($R<0$ and
$Q<-(\frac{27}{4} R^2)^{1/3}$).  (g) and (h) show the
unconditional PDF over the entire $(R,Q)$ plane. Different lines
correspond to different associated eigenvalues: most negative
(dashed), intermediate (solid) and most positive (dashed-dotted).}
\label{fig:RQGeoMin}
\end{figure}

Fig. \ref{fig:RQGeoMin} shows  the PDF of the cosine of the angle
between vorticity and eigendirections with the most negative
(a-b), intermediate (c-d) and most positive (e-f) eigenvalue of
the stress, for both DNS and the model. The different symbols
denote the results obtained in separate quadrants as separated by
the Vieillefosse (i.e. zero discriminant) and the  $R=0$ lines. In
Fig. \ref{fig:RQGeoMin}(g-h) is displayed the unconditional PDF
independent on the quadrant, i.e. as obtained in all regions. As
already observed in \cite{CheMen06}, the model predicts accurately
the preferential alignment with the intermediate eigendirection
(solid line), a trend of being orthogonal to the most contracting
direction (dashed line), and an almost entirely decorrelated trend
with the most extensive eigendirection (dash-dotted line). The
agreement between DNS and the model is excellent in all cases,
even when conditioning on the separate quadrants. It is
interesting to note that in (a) and (b), as well as in (e) and
(f), the alignment PDF is essentially the same in three quadrants
but very different in the $R>0$ and $Q>-(27R^2/4)^{1/3}$ quadrant.
In Fig. \ref{fig:RQGeoMin}(a) and (b) we observe that while the
vorticity is mostly perpendicular to the most contracting
eigendirection, in the top-right quadrant the vorticity is in fact
not orthogonal to the contracting eigen-direction. This is the
``vortex contracting'' quadrant with an unstable focus and one
contracting direction. This would suggest that the vorticity is
aligned with the contracting direction. There is instead no strong
preferred alignment but there is an almost zero probability that
the vorticity is perpendicular to the contracting eigendirection.
But, on average, when taking into account all the possible values
for $R$ and $Q$, (Fig. \ref{fig:RQGeoMin}(g-h)), vorticity remains
weakly orthogonal to this eigendirection. In terms of the
alignments with the intermediate eigendirection, one may have
expected the preferential alignments to come mainly from the
bottom-right quadrant as predicted by the asymptotic diverging
state of the RE equations \cite{Vieille2,Cant1}. Nevertheless, in
Fig. \ref{fig:RQGeoMin}(c) and (d) we observe instead that the
alignment with the intermediate eigendirection occurs quite
independently of the characteristic values in the $(R,Q)$ plane.
In Fig. \ref{fig:RQGeoMin}(e) and (f) we observe that the
alignment with the most extensive strain-rate eigendirection is
random in all quadrant except, again, in the top-right quadrant
where $R>0$ and $Q>-(27R^2/4)^{1/3}$. Vortex ``contraction'', when
it happens, appears to occur because it is mostly orthogonal to
the extensive direction and also `not orthogonal' to the
contracting direction, rather than being preferentially aligned
with the contracting direction. The stochastic model predicts
these non-trivial statistical geometric behaviors quite well.

\section{Pressure Hessian and Viscous term}
\label{sec-flux}

Let us now focus directly on the terms requiring closure, namely
the pressure Hessian and the viscous term, instead of  the
statistics of the velocity gradient tensor considered in the
previous section. One option could be to compare individual
realizations of the model terms with the corresponding DNS values
along Lagrangian trajectories. However, since these terms
fluctuate greatly in the DNS, a statistically more robust
comparison is performed using conditional averages, conditioned on
$R$ and $Q$.

\subsection{Probability current and conditional averages}
\begin{figure}[t]
\center{\epsfig{file=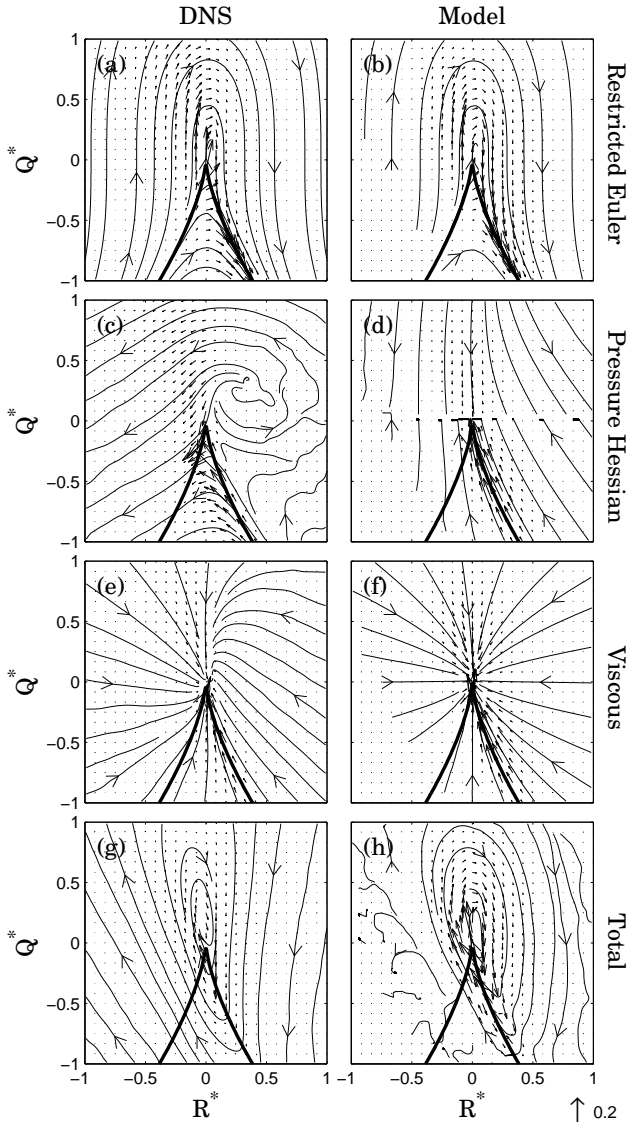,width=8.5cm}} \caption{
Vector and streamline plots of the probability current associated
to the (a-b) RE approximation (Eq. (\ref{eq:WRE})), (c-d) the
pressure Hessian (Eq. (\ref{eq:WP})), (e-f) the viscous term (Eq.
(\ref{eq:WNu})). The total current (Eq. (\ref{eq:WTotal})) is
represented in (g-h).  The scale of the vectors is the same in all
plots and a reference is given below Fig. (h), whose
(non-dimensional) magnitude is $2.10^{-1}$.} \label{fig:RQVec}
\end{figure}

The approach used in Ref. \cite{BosTao02} is followed, based on a
Fokker-Planck equation for the dynamics of $R$ and $Q$. To
summarize the approach, we notice that along a Lagrangian
trajectory, appropriately contracting   (\ref{eq:NS}) with
$\textbf{A}$ and $\textbf{A}^2$, using the Cayley-Hamilton theorem
\cite{Cant1}, one can show that the time evolution of the
invariants $R^*$ and $Q^*$ are given by 
\begin{equation}\label{eq:DynQ}
\frac{dQ^*}{dt^*} = -3R^*-\frac{1}{\sigma^3}A_{ik}H_{ki}^p
-\frac{1}{\sigma^3}A_{ik}H_{ki}^{\nu} ~~\mbox{ and}
\end{equation}
\begin{equation}\label{eq:DynR}
\frac{dR^*}{dt^*} =
\frac{2}{3}\left(Q^*\right)^2-\frac{1}{\sigma^4}A_{ik}A_{kl}H_{li}^p
-\frac{1}{\sigma^4}A_{ik}A_{kl}H_{li}^{\nu} \mbox{ ,}
\end{equation}
where   $\sigma^2 = \langle S_{ij}S_{ij}\rangle$ is the strain variance and $t^*=\sigma
t$ the non-dimensional time. 
Also, $\textbf{H}^p$ stands for (minus) the deviatoric part of the
pressure Hessian, i.e.
\begin{equation}\label{eq:DeviatoricHP}
H_{ij}^p =-\left( \frac{\partial^2p}{\partial x_i\partial x_j
}-\frac{\delta_{ij}}{3}\frac{\partial^2p}{\partial x_k\partial x_k
}\right)\mbox{ ,}
\end{equation}
and $\textbf{H}^\nu = \nu \nabla^2\textbf{A}$ is the viscous term
(recall that in the RE approximation,
$\textbf{H}^p=\textbf{H}^\nu=0$). The Fokker-Planck equation
describing the time evolution of the joint density $\mathcal
P(Q^*,R^*)$ may be written as \cite{Ris84}:
\begin{equation}\label{eq:FokkerRQ}
\frac{\partial \mathcal P}{\partial t^*} + \begin{pmatrix}
  \frac{\partial }{\partial Q^*} \\
  \frac{\partial }{\partial R^*}
\end{pmatrix}.\vct{\mathcal W}=0\mbox{ ,}
\end{equation}
where the divergence of the \textit{probability current}
$\vct{\mathcal W}$ controls time variations of the joint
probability density $\mathcal P$. The probability current can be
written in terms of conditional averages as
\begin{equation}\label{eq:WTotal}
\vct{\mathcal W} =\left\langle
\begin{pmatrix}
  \frac{dQ^*}{dt^*} \\
  \frac{dR^*}{dt^*}
\end{pmatrix}\bigg | Q^*,R^*\right\rangle \mathcal P(Q^*,R^*).
\end{equation}
It can be decomposed into $\vct{\mathcal W} = \vct{\mathcal
W}_{RE}+\vct{\mathcal W}_p+\vct{\mathcal W}_\nu$,   with
\begin{equation}\label{eq:WRE}
\vct{\mathcal W}_{RE} =\left\langle
\begin{pmatrix}
  -3R^* \\
  \frac{2}{3}\left(Q^*\right)^2
\end{pmatrix} \bigg | Q^*,R^*\right\rangle\mathcal P(Q^*,R^*)
\end{equation}
which describes the deterministic (closed) part of the evolution of the two
invariants,
\begin{equation}\label{eq:WP}
\vct{\mathcal W}_{p} =\left\langle
\begin{pmatrix}
  -A_{ik}H_{ki}^p/\sigma^3 \\
  -A_{ik}A_{kl}H_{li}^p/\sigma^4
\end{pmatrix} \bigg | Q^*,R^*\right\rangle\mathcal P(Q^*,R^*)\mbox{ ,}
\end{equation}
describing the pressure Hessian effects on the evolution of $R^*$
and $Q^*$ and finally,
\begin{equation}\label{eq:WNu}
\vct{\mathcal W}_\nu =\left\langle
\begin{pmatrix}
  -A_{ik}H_{ki}^\nu/\sigma^3 \\
  -A_{ik}A_{kl}H_{li}^\nu/\sigma^4
\end{pmatrix}\bigg | Q^*,R^*\right\rangle\mathcal P(Q^*,R^*)\mbox{ ,}
\end{equation}
describing the effects of the viscous term. An additional current might be considered in this description, linked to an additional forcing term that has been neglected in the Navier-Stokes equations (Eq. (\ref{eq:NS})). This forcing is indeed negligible in front of the other terms of the rhs. of Eq. (\ref{eq:NS}) since it can be written as the (small-scale) gradient of the large-scale forcing of the velocity, and thus, we will neglect its associated probability current.

Conversely, in the Fokker-Planck equation (Eq. (\ref{eq:FokkerRQ}))  for the joint probability distribution of $R^*$ and $Q^*$
obtained from the model (Eq. (\ref{eq:Determourmodel})), one has to take into account another term which comes from
the delta-correlated Gaussian forcing. Appendix \ref{app:B}  provides the required
background needed to compute the probability flux resulting from the
stochastic forcing term in our model, i.e. the diffusion 
terms entering in Eqs. (\ref{eq:DynQ}) and (\ref{eq:DynR}). It is shown that the currents associated to the deterministic and random parts of the joint stochastic evolution of $R$ and $Q$ predicted by the model (see Eq. (\ref{eq:SysRQSDE})) are of the same order of magnitude.

\subsection{Results}
In Fig. \ref{fig:RQVec} the vector plots and associated
streamlines corresponding to the various probability flux terms
are presented. Both results obtained from DNS and from the model
are shown.

First, as reference we present in Fig. \ref{fig:RQVec}(a-b) the
closed RE current $\vct{\mathcal W}_{RE}$ (Eq. (\ref{eq:WRE})). As
is well-known \cite{Vieille2,Cant1}, the deterministic
$\vct{\mathcal W}_{RE}$ probability current pushes probabilities
towards the right tail of the Vieillefosse line. Since the model
predicts acurately the joint probability $\mathcal P(Q^*,R^*)$,
agreement between DNS and the model predictions (length of
vectors) is quite good because the self-streching term $-\textbf{A}^2$ is taken into account exactly in the model (Eq. \ref{eq:Determourmodel}) .

The action of the pressure Hessian, given by the probability
current $\vct{\mathcal W}_p$ and shown in Fig.
\ref{fig:RQVec}(c-d), is quite interesting. From the DNS data,
two main pressure Hessian effects can be observed. First, the
pressure Hessian counteracts the effects induced by the RE terms
since the flux goes towards the center of the RQ plane along the
right tail Vieillefosse line. This feature is well reproduced by
the model, with vector magnitudes of the same order. Another
important effect of the pressure Hessian is that in the $R<0$ left
half-plane, the probability current  leads the probability towards
the left tail of the Vieillefosse line, namely towards
dissipation-enstrophy production dominated region (in lower-left
direction). This feature is \textit{not} reproduced by the model,
which instead appears to act exclusively in vertical direction,
upward in the $Q<0$ plane, and downward in the $Q>0$ side. This
explains perhaps why the model leads to an underestimation (see
Fig. \ref{fig:RQ}) of the probability of dissipation-enstrophy
production events (i.e. the bottom-left quadrant). At the same
time, the absence of ``left-ward'' flux out of the
enstrophy-dissipation production region  (i.e. top-right quadrant)
may explain why the model overpredicts the probability of events
in that quadrant. A very marked feature of the DNS results is that
the magnitudes of the vectors are essentially negligible in the
entire vortex contraction quadrant above the right Vieillefosse
line ($R>0$ and $Q>-(27R^2/4)^{1/3}$), leading to some uncertainty
in the computed streamlines there.

Another main difference between DNS and model predictions is the
fact that for the model, $\vct{\mathcal W}_p$ vanishes at
vanishing $Q$, but \textit{not} for the DNS. A general feature of
the pressure Hessian model is that its deviatoric part is directly
proportionnal to $\mbox{Tr}(\textbf{A}^2)= -2Q$ (see Eq.
(\ref{eq:modHess})). Incidentally, the same occurs in further
generalizations that have been proposed by Gibbon and Holm
\cite{GibHol07} (see their Eq. (5.8)), namely
\begin{equation}\label{eq:GibHolModel}
\vct{\nabla}\vct{\nabla}p = -\left[ \sum_{n=1}^N c_n
\frac{\textbf{G}_\textbf{n}}{\mbox{Tr}(\textbf{G}_\textbf{n})}\right]
\mbox{Tr}(\textbf{A}^2) \mbox{ with }  \sum_{n=1}^N c_n =1
\end{equation}
where the scalars $c_N$ are undertermined and $\textbf{G}_\textbf{n}$
are any non-singular symmetric tensors. Once again, we can see that
the deviatoric part of the pressure Hessian ${\textbf{H}^p}$ is still
proportional to $\mbox{Tr}(\textbf{A}^2)$. 
For the sake of completeness, let us remark that at least formally, this 
issue does not arise in the matrix shell model of \cite{BifChe07}. This is
because the non-local closure terms in the matrix shell model \cite{BifChe07} are not 
directly proportional to $\mbox{Tr}(\textbf{A}^2)$ since the connection 
to pressure Hessian and Poisson equation for pressure is not included in that approach. 
It would be very interesting to check if the comparison with
DNS for the equivalent probability current in the matrix shell model, i.e. based on
relevant portions of the quadratic nonlinear interaction terms,  is better or not. 
Such studies are left for future work.

\begin{figure}[t]
\center{\epsfig{file=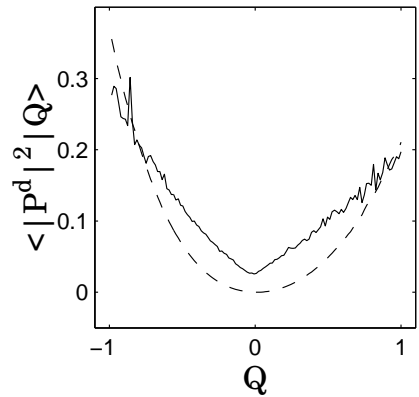,width=5.5cm}}
\caption{Conditional  average of the norm square of the deviatoric
part of the pressure Hessian,  $\left \langle
\left|{\textbf{H}^p}\right|^2 | Q \right\rangle$, with respect to
the invariant $Q$. Both $ \left|{\textbf{H}^p}\right|^2$ and $Q$,
are non-dimensionalized by their respective standard deviations.
DNS results (solid line) and model predictions (dashed line) are
shown.} \label{fig:CondHPonQ}
\end{figure}

To more clearly isolate the behavior of the pressure Hessian near $Q=0$ line, we study the
magnitude of the anisotropic (i.e. deviatoric) part of the pressure
Hessian, conditioned on the local value of $Q$. Fig.
\ref{fig:CondHPonQ} shows the conditional average of the norm (square)
of the deviatoric part of the pressure Hessian, i.e.  $\left \langle
  \left|{\textbf{H}^p}\right|^2 | Q \right\rangle$, where
$\left|{\textbf{H}^p}\right|^2 = \mbox{Tr}\left( {\textbf{H}^p}
  \left({\textbf{H}^p}\right)^\top\right)$, as a function of the local
value of the invariant $Q$, for both the DNS and the model. For
vanishing $Q$, the conditional average $|{\textbf{H}^p}|^2$ from the
DNS does not vanish. As discussed before this property is not
reproduced by the existing models for the pressure Hessian, namely the
tetrad model (Eq.  (\ref{eq:ModHessTetrad})), the CM06 model
(Eq. (\ref{eq:modHess})) and the generalized tetrad model
(Eq. (\ref{eq:GibHolModel})) because all of them predict a pressure
Hessian proportional to $Q$.  Secondly, one can see that for the range
of $Q$ under consideration (i.e. $Q\in [-\sigma_Q,\sigma_Q]$
($\sigma_Q$ stands for the standard deviation of $Q$), the conditional
average of $\left|{\textbf{H}^p}\right|^2$ behaves as $ \sim |Q|$ (up
to a positive additive constant) whereas the model predicts it to be
proportional to $Q^2$. The fact that the model predicts a quadratic
behavior can be understood from a Taylor's development, namely
${\textbf{H}^p}\approx -Q\tau_K\textbf{S}$ leading to $\left \langle
  \left|{\textbf{H}^p}\right|^2 | Q \right\rangle \sim Q^2$ since
$\langle\tau_K^2\left|\textbf{S}\right|^2\rangle\sim 1$.  The small asymmetry in the
quadratic behavior seen in Fig.  \ref{fig:CondHPonQ} is caused by
higher order terms entering in the expansion for the model.

In terms of the viscous term, one observes in Fig.
\ref{fig:RQVec}(e-f) that  the model reproduces the probability
flux reasonably well. Consistent with the observations already
made in \cite{BosTao02} the viscous effect is to push the
probabilities towards vanishing $R$ and $Q$, not only along the
Vieillefosse line, but everywhere. We notice that the model
overpredicts the magnitudes, i.e. at this Reynolds number the
model provides too strong damping but is qualitatively correct.

In  Fig. \ref{fig:RQVec}(g-h)  is shown the sum of all these
terms, namely the total probability current $\vct{\mathcal W} =
\vct{\mathcal W}_{RE}+\vct{\mathcal W}_p+\vct{\mathcal W}_\nu$.
For the model case, another term coming from the Gaussian
delta-correlated forcing (see Appendix \ref{app:B} and Eq.
(\ref{eq:WideProbCurrMod})) has been added. The circular motion
around the origin of the RQ plane has already been reported in
Refs. \cite{ooimar99,ChePum99}. 
At this point, from Fig.  \ref{fig:RQVec} one can observe that all the terms (self-stretching, pressure Hessian and viscous Laplacian) entering in the Navier-Stokes equations (Eq. (\ref{eq:NS})) are of the same order of magnitude (viscous Laplacian is a little bit smaller than the two other terms, but not by much). Similar conclusions can be drawn for the deterministic terms entering in the model (Eq. (\ref{eq:Determourmodel}) and figs. \ref{fig:RQVec}(b-d-f)), although, as  shown in appendix \ref{app:B}, the amplitude of the forcing is not negligible either.  Focusing on the total probability current  \ref{fig:RQVec}(g-h), we reach the conclusion that the fact that the modeled pressure Hessian (fig. \ref{fig:RQVec}(d)) is not able to reproduce the probability flux towards $R^*<0\mbox{ and }Q^*<0$ regions as it is observed in DNS (fig. \ref{fig:RQVec}(c)), explains why the model over predicts  $R^*>0\mbox{ and }Q^*>0$ regions and under predicts $R^*<0\mbox{ and }Q^*<0$ regions, as observed in Fig. \ref{fig:RQ}. We also remark that streamlines shown in the left part of Fig. \ref{fig:RQVec}(h) are not significant because of the very low values of the joint probability $\mathcal P (Q^*,R^*)$.

\begin{figure}
\center{\epsfig{file=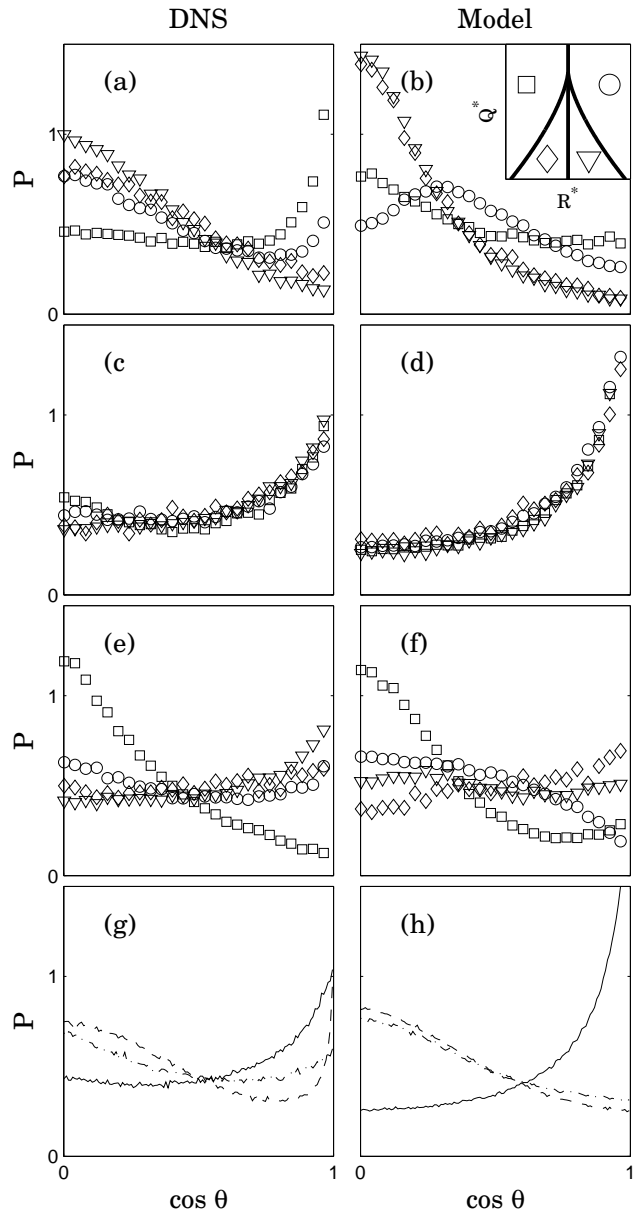,width=8.5cm}} \caption{PDFs
of the cosine of the angle between vorticty and the different
eigendirections of pressure Hessian tensor ${\textbf{H}^p}$: (a)
and (b) for the smallest eigenvalue eigendirection, (c) and (d)
for the intermediate eigendirection, and (e) and (f) for the most
positive eigen-direction. As it is schematically displayed in the inset of Fig. \ref{fig:RQGeoHP}(b), in a similar fashion as in Fig. \ref{fig:RQGeoMin}, different symbols are
obtained from the four different regions of the $(R,Q)$ plane delimited by the $R>0$ and the \textit{Vieillefosse} (or zero-discriminant) lines given by $Q=-(\frac{27}{4} R^2)^{1/3}$: $\circ$ ($R>0$ and
$Q>-(\frac{27}{4} R^2)^{1/3}$), $\triangledown$ ($R>0$ and
$Q<-(\frac{27}{4} R^2)^{1/3}$), $\square$ ($R<0$ and
$Q>-(\frac{27}{4} R^2)^{1/3}$), $\lozenge$ ($R<0$ and
$Q<-(\frac{27}{4} R^2)^{1/3}$).  Figs (g) and (h) show the
unconditional PDF over the entire $(R,Q)$ plane. Different lines
correspond to different associated eigenvalues: most negative
(dashed), intermediate (solid) and most positive (dashed-dotted).} \label{fig:RQGeoHP}
\end{figure}

\section{Vorticity alignments with pressure Hessian and viscous Laplacian eigendirections}
\label{sec-alignments}

\subsection{Pressure Hessian}

Here we focus on vorticity alignment properties along the
eigendirections of the pressure Hessian. It has been
derived, in the inviscid limit (Euler equations)
\cite{Ohk93,GalGib97,GibHol06} that vorticity $\omega_i$ tends to
be simultaneously an eigenvector of the rate of strain tensor
$\textbf{S}$ and the pressure Hessian. When $\nu=0$ (see Eq.
(\ref{eq:EnstrEvol})),
\begin{equation}
\frac{d\omega_i}{dt} = S_{ij}\omega_j\mbox{ ,}
\end{equation}
and taking another time derivative and using the time evolution of
$\textbf{S}$ (see Eq. (\ref{eq:DissEvol})), we get
\begin{equation}
\frac{d^2\omega_i}{dt^2} = -\frac{\partial ^2p}{\partial
x_i\partial x_j }\omega_j\mbox{ .}
\end{equation}
Following Ref. \cite{Ohk93}, we then notice that \textit{if
vorticity of a fluid particle continues to be an eigenvector of
the rate-of-strain tensor, then it is also an eigenvector of the
pressure Hessian}. To see if such a trend  is observed  in a
finite viscosity turbulent flow we will quantify the aligments of
vorticity with the eigendirections of the pressure Hessian. Such
an analysis based on DNS has been already performed \cite{TsinoNice,Kal06},
but here the purpose is to compare results with predictions of the
model.

Let us focus on alignment properties of vorticity with respect to
the eigendirections of the deviatoric part of the pressure Hessian
(i.e. $-H_{ij}^p$ defined in Eq. (\ref{eq:DeviatoricHP})).
Alignment PDFs are shown in Fig. \ref{fig:RQGeoHP}, presented in a
similar fashion as in Fig. \ref{fig:RQGeoMin}. We can see in Fig.
\ref{fig:RQGeoHP}(a) that vorticity is preferentially orthogonal
to the eigendirection of the smallest eigenvalue, except in the
top-left quadrant (i.e. $R<0$ and $Q>-(27/4R^2)^{1/3}$) where the
local topology is dominated by one direction of streching and a
stable focus. Also in this quadrant, vorticity is preferentially
aligned with the extending eigendirection. The model predicts a
slightly different picture since vorticity is predicted to be also
preferentially othogonal to the eigendirection except in the
top-right quadrant (i.e. $R>0$ and $Q>-(27/4R^2)^{1/3}$) for which
local topology is dominated by one compressive direction and an
unstable focus.  

In Fig. \ref{fig:RQGeoHP}(c-d), we focus on the
pressure  Hessian eigendirection of its intermediate eigenvalue.
In a similar way as with eigendirections of the rate-of-strain,
vorticity is preferentially aligned with this eigendirection in
all the quadrants, and this is also very well predicted by the
model. About the eigendirection of the largest eigenvalue, we can
see that for the DNS (Fig. \ref{fig:RQGeoHP}(e)) vorticity is
weakly preferentially aligned with the eigendirection, except
again in the top-left quadrant where vorticity is clearly
preferentially orthogonal to this eigendirection. In Fig.
\ref{fig:RQGeoHP}(f), we can see that the model predicts most of
the trends quite well in all the quadrants. For the average
results over all quadrants, it can be seen in
 Figs. \ref{fig:RQGeoHP}(g-h)  that the
model predicts with a fairly good accuracy the behavior of
vorticity with the eigendirection of the intermediate eigenvalue
(although it overpredicts the peak a little bit). In the other
extremal eigendirections, the model reproduces the moderate peak
at $\cos(\theta)\sim 0$ but misses the narrow peaks near alignment
at $\cos(\theta)\sim 1$.

The fact that the model reproduces very well the events for which
vorticity happens to be an eigenvector of the
rate-of-strain tensor can be understood phenomenologically in the
following way. When vorticity is an eigenvector of the
rate-of-strain, i.e. $\textbf{S}\vct{\omega} = \beta
\vct{\omega}$, then vorticity is also an eigenvector of the
velocity gradient tensor itself, namely $\textbf{A}\vct{\omega} =
(\textbf{S}+\vct{\Omega})\vct{\omega} = \beta\vct{\omega}$, since
by definition $\vct{\Omega}\vct{\omega}=0$. Let us notice that
vorticity is also an eigenvector of $\textbf{A}^\top=
\textbf{S}-\vct{\Omega}$ with the same eigenvalue $\beta$. It is
then straightforward to show by induction that for any power $n\in
\mathbb N$, $\textbf{A}^n\vct{\omega}=(\textbf{A}^\top)^n\vct{\omega}
= \beta^n\vct{\omega}$. For these very particular events in which
vorticity is an eigenvector of the rate-of-strain, one notices
than the matrix exponential entering the (inverse) recent
Cauchy-Green tensor (in Eq. (\ref{eq:MyCG})) can be
written as
\begin{equation}\label{eq:ExpCG}
\textbf{C}_{\tau_K}^{-1} = e^{-\tau_K \textbf{A}^\top}e^{-\tau_K
\textbf{A}}=\sum_{n,m=0}^
{+\infty}\frac{(-\tau_K)^{n+m}}{n!m!}(\textbf{A}^\top)^n\textbf{A}^m
\end{equation}

From Eq. (\ref{eq:ExpCG}) it is easily seen that vorticity is also
an eigenvector of both the recent Cauchy-Green tensor (Eq.
(\ref{eq:MyCG})) and its inverse. For example,
$\textbf{C}_{\tau_K}^{-1}\vct{\omega} =
e^{-2\tau_K\beta}\vct{\omega}$. Finally, since the pressure
Hessian is modeled as proportional to $\textbf{C}_{\tau_K}^{-1}$
(Eq. (\ref{eq:modHess})), we can state here that \textit{the
present model is such that when the vorticity is an eigenvector of
the rate-of-strain, then it is also an eigenvector of the modeled
pressure Hessian, the ordering of the associated eigenvalues being
respected in absolute value}. More precisely, if vorticity is an
eigenvector of $\textbf{S}$ with respective eigenvalue $\beta$,
then vorticity is also an eigenvector of the pressure Hessian with
eigenvalue
$-\frac{\mbox{Tr}(\textbf{A}^2)}{\mbox{Tr}(\textbf{C}_{\tau_K}^{-1})}
e^{-2\tau_K\beta}$.

\begin{figure}[t]
\center{\epsfig{file=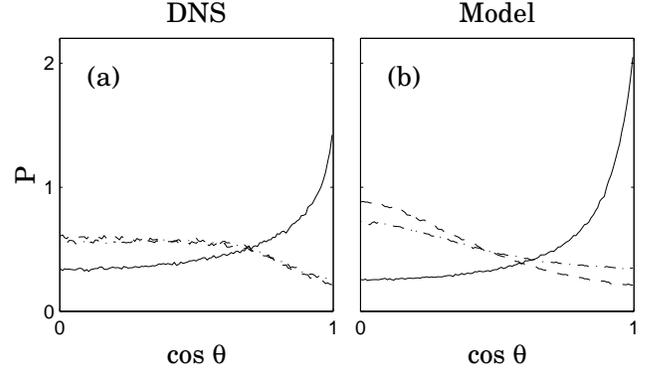,width=8.5cm}}
\caption{PDF of the cosine between vorticity and eigendirections
of $\nu\Delta\textbf{S}$ associated to the smallest (dashed line),
intermediate (solid line) and biggest (dot-dashed line)
eigenvalues of the strain-rate tensor.}
\label{fig:AlignVortViscous}
\end{figure}

\subsection{Viscous Term}

Let us now focus on the geometrical properties of the viscous
tensor, namely $\nu\nabla^2\textbf{A}$, appearing in Eq.
(\ref{eq:NS}). Let us begin with its symetric part
$\nu\nabla^2\textbf{S}$. The present model (i.e. Eq.
(\ref{eq:Determourmodel})) contains the closure for the viscous
term pointed  of Eq. (\ref{eq:ModViscA}) that stated that the
Laplacian of $\textbf{A}$ is  proportionnal to $\textbf{A}$
itself. In terms of eigendirections, it is assumed that both
$\textbf{S}$ and $\nabla^2\textbf{S}$ have the same
eigendirections. Among others, alignment properties of vorticity
$\vct{\omega}$ and eigendirections of $\nabla^2\textbf{S}$ should
be exactly the same as alignments of vorticity with
eigendirections of $\textbf{S}$. To determine whether this is
observed in DNS flows, we present in Fig.
\ref{fig:AlignVortViscous} PDFs of the cosine of the angle between
vorticity and eigendirections of the viscous term, for both DNS
(a) and the model (b). Fig. \ref{fig:AlignVortViscous}(b) is in
fact the same as Fig. \ref{fig:RQGeoMin}(h) and is reproduced here
for convenience. We see clearly that the overall geometrical
picture are really close between DNS and the model, and that there
is preferential alignment of vorticity with the eigendirection
associated to the intermediate eigenvalue of the Laplacian term as
well.

\begin{figure}[t]
\center{\epsfig{file=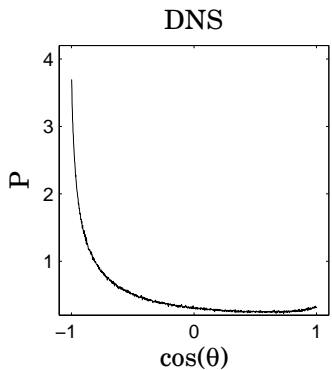,width=4.5cm}}
\caption{PDF of the cosine of the angle $\theta$ between vorticity
$\omega_i = -\frac{1}{2}\varepsilon_{ijk}\Omega_{jk}$ and $\vct{\varpi}$, the
Laplacian of vorticity vector  $\varpi_i = -\frac{1}{2}\varepsilon_{ijk}\nu\nabla^2\Omega_{jk}$, obtained
from DNS.}
\label{fig:AlignVortVisc}
\end{figure}

Let us now focus on the antisymmetric part of the viscous term,
namely $\nu\nabla^2\vct{\Omega}$. 
The vorticity vector is given by   $\omega_i =
-\frac{1}{2}\varepsilon_{ijk}\Omega_{jk}$. The viscous term
 $\vct{\varpi}=\nu\nabla^2\vct{\omega}$ can also be written as a vector, i.e. $\varpi_i =
-\frac{1}{2}\varepsilon_{ijk}\nu\nabla^2\Omega_{jk}$.  The model for the
viscous term (Eq. (\ref{eq:RFDModViscA})) implies that the angle
$\theta=(\vct{\varpi},\vct{\omega})$ between the vorticity and the vorticity
Laplacian is fixed and equals $\pi$,
i.e. same direction but opposite orientation, since the model (Eq.
(\ref{eq:RFDModViscA})) is proportional to the velocity gradient
tensor $\textbf{A}$ with a negative coefficient. One may wonder if
this is consistent with DNS data. We represent in Fig.
\ref{fig:AlignVortVisc} the PDF of $\cos \theta$ estimated from
the same DNS fields. Clearly we see that the two vectors share
preferentially the same direction, but opposite orientation. The model is therefore 
consistent with the observed alignment trends of the full Laplacian of velocity gradient. 

\section{Conclusions}
\label{sec-conclusions}

Extensive comparisons have been made between predictions of a new
stochastic Lagrangian model for the velocity gradient tensor and
results from DNS at a corresponding moderate Reynolds number.  The
model reproduces many inherent geometric and statistical
properties of small-scale turbulence quite well.  The statistics
of alignment angles between vorticity and the principal axes of
the rate-of-strain tensor are very well reproduced.  The joint
statistics of velocity gradient invariants $R$ and $Q$ are also
reproduced well. Specifically, the joint PDF's elongation into the
top-left and bottom-right quadrants observed in the DNS also
occurs in the model. Some differences occur in the model in the
top-right and bottom-left quadrants. In order to directly assess
the action of the modeled pressure and viscous terms in a
statistically robust fashion that takes into account the local
topology of the flow,  the probability current $\vct{\mathcal W}$ has
been studied.  The agreement between DNS and model predictions
is good near the dominant Vieillefosse tail in the lower-right
quadrant of the $(R,Q)$ plane. However, in the
dissipation-enstrophy production dominated region (bottom-left),
the model does not reproduce the true dynamics and requires
further developments. Finally,  the alignment properties of
vorticity with respect to the principal axes of the pressure
Hessian tensor have been studied. The model reproduces quite well
the preferential alignment of vorticity with the eigendirection
associated to the intermediate eigenvalue. We elucidate the fact
that in the model an eigenvector of the rate-of-strain is also an
eigenvector of the pressure Hessian, and this is in fact
consistent with known behavior of vorticity in the inviscid limit
(i.e. the Euler equations).

This analysis has confirmed that the stochastic model is capable
of predicting many non-trivial properties of small-scale
turbulence as described by the geometric and statistical
properties of the velocity gradient tensor. Nevertheless, there
appear to be difficulties in specific regions of the flow,
especially those in which the vorticity is being contracted such
as in the top-right or bottom-left portion of the invariant $(R,Q)$
plane. Whether these drawbacks of the model are also related to
the difficulties observed when raising the Reynolds number of the
flow also remains to be explored.

Acknowledgements:  We thank  S. Chen, R. Ch\'etrite, G. Eyink, K.
Gawedzki and Y. Li for useful discussions. L.C is supported by
postdoctoral Fellowship from the Keck Fundation and C.M. by the
National Science Foundation.

\appendix

\section{Definition and implementation of tensor Gaussian forcing}
\label{ann:DefBruit} In this appendix,   the
tensorial Gaussian forcing $d\textbf{W}$ entering in the 
model (Eq. (\ref{eq:Determourmodel})) is described. It can be written as
\begin{equation}
dW_{ij} = D_{ijkl}dB_{kl}
\end{equation}
where $D_{ijkl}$ are the diffusion coefficients and $d\textbf{B}$
is a tensorial isotropic Wiener process, whose components are such
that
\begin{equation}
\langle dB_{ij}\rangle=0 \mbox{ and } \langle
dB_{ij}dB_{kl}\rangle=2dt \delta_{ik}\delta_{jl}\mbox{ .}
\end{equation}
The coefficients $D_{ijkl}$ are chosen such that the noise
$d\textbf{W}$ is consistent with a trace-free, homogeneous and
isotropic tensor, of a given unit variance, namely $\langle
dW_{ij}dW_{kl}\rangle=2dtD_{ijpq}D_{klpq}$ with
\begin{equation}\label{eq:Chol}
D_{ijpq}D_{klpq}
=2\delta_{ik}\delta_{jl}-\frac{1}{2}\delta_{ij}\delta_{kl}-
\frac{1}{2}\delta_{il}\delta_{jk}\mbox{ .}
\end{equation}
As a consequency, longitudinal components of the noise
$d\textbf{W}$ are of variance $2dt$, and $4dt$ for the transverse
ones. Also, let us recall that the dimension of the diffusion
coefficients is $\mbox{time}^{-3/2}$.  If the tensor ${\bf D}$ is assumed isotropic itself, then the unique solution of Eq. (\ref{eq:Chol})  is given by
\begin{equation}\label{eq:SolIsotropic}
D_{ijpq}
=a\delta_{ij}\delta_{pq}+b\delta_{ip}\delta_{jq}
+c\delta_{iq}\delta_{jp}\mbox{ ,}
\end{equation}
with
\begin{equation}
a
=\frac{1}{3}\frac{3+\sqrt{15}}{\sqrt{10}+\sqrt{6}}; b=-\frac{\sqrt{10}+\sqrt{6}}{4}; c= \frac{1}{\sqrt{10}+\sqrt{6}}\mbox{ .}
\end{equation}

\section{Stochastic differential equations}
\label{app:B} 

The basic tensorial stochastic differential equation (SDE) Eq.
(\ref{eq:Determourmodel}) can be written as
\begin{equation}\label{eq:ItoStoch}
dA_{ij} = V_{ij}dt + D_{ijkl}dB_{kl} \mbox{ ,}
\end{equation}
where $V_{ij}$ is the drift coefficients representing the
self-streching, pressure Hessian and viscous terms entering in Eq.
(\ref{eq:Determourmodel}) and $D_{ijkl}dB_{kl}$ is the forcing term described in
Appendix \ref{ann:DefBruit}. The associated Fokker-Planck equation for the joint
probability of the velocity gradients $f (\textbf{A};t)$, of
is given by
\begin{equation}\label{eq:FokkItoStoch}
\frac{\partial f}{\partial t} = -\frac{\partial}{\partial
A_{ij}}\left[f V_{ij} \right]+ \frac{\partial ^2 }{\partial
A_{ij}\partial A_{kl}}\left[ fD_{ijpq}D_{klpq}\right]\mbox{ .}
\end{equation}
\begin{figure}[t]
\center{\epsfig{file=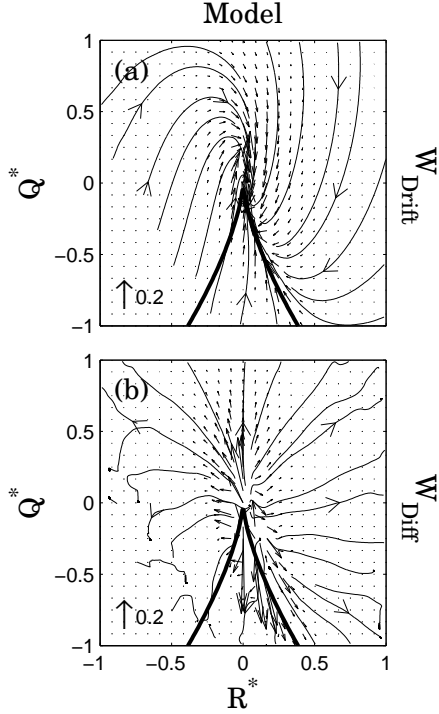,width=6cm}} \caption{Vector and
streamline plots of the probability current for the model
associated to the (a) the drift coefficients and (b) the diffusion
coefficients (see text and Eq. (\ref{eq:WideProbCurrMod})). The
scale of the vectors is the same, reference is given in
each figures, whose (un-dimensionalized) magnitude is
$2.10^{-1}$.} \label{fig:DriDifMod}
\end{figure}

Here we are interested in the joint probability
$\mathcal P(Q,R)$ of two invariants of the velocity gradients,
namely $R$ and $Q$ (c.f. Eq. (\ref{eq:FokkerRQ})). In It\^o
interpretation, both the stochastic equations governing the
dynamics of $R$ and $Q$  and the associated Fokker-Planck equation
can be computed (c.f. \cite{GirPop90,Oks03,Ris84,KloPla92}) from
the evolution of $\textbf{A}$ (Eq. (\ref{eq:ItoStoch})). To do so,
one needs to know how a stochastic differential equation is
written under a non-linear transformation since $R$ and $Q$ are
nonlinear functions of the components of $\textbf{A}$.

In general terms, let us call such a time dependent nonlinear
transformation $\vct{\xi}(t,\textbf{A}): \textbf{A}\mapsto
\vct{\xi}(t,\textbf{A})$ with components $\xi_k$, $k\in
\{1,2,...,N\}$. Starting from the SDE of \textbf{A} (Eq.
(\ref{eq:ItoStoch})), general formula
\cite{GirPop90,Oks03,Ris84,KloPla92} give the new stochastic
differential equations that governes $\vct{\xi}$, namely
\begin{align}\label{eq:GeneChangVar}
d\vct{\xi} = &\left[ \frac{\partial \vct{\xi}}{\partial
t}+\frac{\partial \vct{\xi}}{\partial A_{ij}}V_{ij} + \frac{1}{2}
\frac{\partial^2 \vct{\xi}}{\partial A_{ij}\partial
A_{pq}} D_{ijrs}D_{pqrs} \right]dt \notag \\
&+ \frac{\partial \vct{\xi}}{\partial A_{ij}}D_{ijrs}dB_{rs}
\end{align}

To compute the time evolution of the invariants $R$ and $Q$, we
use Eq. (\ref{eq:GeneChangVar}) with the particular (time
independent) nonlinear transformation $\xi_1 = Q =
-\mbox{Tr}(\textbf{A}^2)/2= -A_{ij}A_{ji}/2 $ and $\xi_2 = R =
-\mbox{Tr}(\textbf{A}^3)/3=-A_{ij}A_{jk}A_{ki}/3$. We get
\begin{equation}\label{eq:SysRQSDE}
\left\{
\begin{array}{ll}
dQ &= \left[-V_{ij}A_{ji}-\frac{1}{2}D_{ijpq}D_{jipq}\right]dt \\
&-A_{ji}D_{ijpq}dB_{pq}\\
dR &= \left[-V_{ij}A_{jq}A_{qi}-A_{li}D_{ijpq}D_{jlpq}\right]dt\\
&-A_{jr}A_{ri}D_{ijpq}dB_{pq}
\end{array}
\right.\mbox{ ,}
\end{equation}
where, using former notations, we notice that $-V_{ij}A_{ji} =
-3R-H^p_{ij}A_{ji}-H^\nu_{ij}A_{ji}$ and using the Cayley-Hamilton
theorem, $-V_{ij}A_{jq}A_{qi} =
\frac{2}{3}Q^2-H^p_{ij}A_{jq}A_{qi}-H^\nu_{ij}A_{jq}A_{qi}$.
Furthermore, the ``spurious" drift terms coming from the
delta-correlated Gaussian noise vanish, i.e. using Eq.
(\ref{eq:Chol}) one can show that $D_{ijpq}D_{jipq}=0$ and
$A_{li}D_{ijpq}D_{jlpq}=0$. From a straightforward manner
\cite{GirPop90,Oks03,Ris84,KloPla92}, we get from Eq.
(\ref{eq:SysRQSDE}) the corresponding Fokker-Planck equation for the
joint probability $\mathcal P(Q,R)$
\begin{equation}\label{eq:FirstFPeq}
\frac{\partial \mathcal P}{\partial t} = -\frac{\partial}{\partial
\xi_i }[\mathcal P N_i] +\frac{\partial}{\partial \xi_i\partial
\xi_j }[\mathcal P M_{ij}]\mbox{ ,}
\end{equation}
In Eq. (\ref{eq:FirstFPeq}), the new coefficients $N_i$ and $M_{ij}$ can be easily obtained from Eq. (\ref{eq:SysRQSDE}), although, they are not known as functions of $\xi_1=Q$ and $\xi_2=R$. Therefore, we will use conditional averages to estimate them. Henceforth, we will write the Fokker-Planck equation with conditional averages and obtain 
\begin{equation}
\frac{\partial \mathcal P}{\partial t} = -\frac{\partial}{\partial
\xi_i }\langle \mathcal P N_i \vert  \vct{\xi}\rangle+ \frac{\partial}{\partial \xi_i\partial
\xi_j }\langle \mathcal P M_{ij}  \vert \vct{\xi} \rangle  \mbox{ ,}
\end{equation}
where the drift coefficients $N_i$ are given by
\begin{equation}
N_i = \begin{pmatrix}
  -3R-H^p_{ij}A_{ji}-H^\nu_{ij}A_{ji} \\ \frac{2}{3}Q^2-H^p_{ij}A_{jq}A_{qi}-H^\nu_{ij}A_{jq}A_{qi}
\end{pmatrix}
\end{equation}
and the diffusion elements by
\begin{align*}
M_{11}&=2\mbox{Tr}\left( \textbf{A}\textbf{A}^\top\right) +Q\mbox{ ,}\\
M_{12}&=M_{21}=2\mbox{Tr}\left(\textbf{A}^\top\textbf{A}^2\right)+\frac{3}{2}R \mbox{ ,}\\
M_{22}&=2\mbox{Tr}\left((\textbf{A}^\top)^2\textbf{A}^2\right)-2Q^2-\frac{1}{2}\mbox{Tr}\left( \textbf{A}^4\right)
\end{align*}
Finaly, using again the general transformation Eq.
(\ref{eq:GeneChangVar}), the probability current $\vct{\mathcal W}$
of the joint probability $\mathcal P (Q^*,R^*)$ of the
non-dimensional invariants $Q^*$ and $R^*$, entering in the non-dimensional Fokker-Planck equation \begin{equation}
\frac{\partial \mathcal P}{\partial t^*} + \begin{pmatrix}
  \frac{\partial }{\partial Q^*} \\
  \frac{\partial }{\partial R^*}
\end{pmatrix}.\vct{\mathcal W}=0\mbox{ ,}
\end{equation}
 is given by
$\vct{\mathcal W}=\vct{\mathcal W}_{\mbox{drift}}+\vct{\mathcal
W}_{\mbox{diff}}$ with (we recall that $\sigma^2 = \langle
S_{ij}S_{ij}\rangle$)
\begin{widetext}
\begin{equation}\label{eq:WideProbCurrMod}
\vct{\mathcal W} =\overbrace{\left\langle
\begin{pmatrix}
  N_1/\sigma^3 \\
  N_2/\sigma^4
\end{pmatrix}\bigg | Q^*,R^*\right\rangle \mathcal P(Q^*,R^*)}^{\vct{\mathcal W}_{\mbox{drift}}}
\underbrace{- \begin{pmatrix}
  \frac{\partial}{\partial Q^*} \frac{\partial}{\partial
  R^*}
\end{pmatrix}\left[\left\langle\begin{pmatrix}
  M_{11}/\sigma^5 & M_{12}/\sigma^6 \\
  M_{21}/\sigma^6 & M_{22}/\sigma^7
\end{pmatrix}\bigg |  Q^*,R^*\right\rangle \mathcal
P(Q^*,R^*)\right]}_{\vct{\mathcal W}_{\mbox{diff}}}\mbox{ .}
\end{equation}
\end{widetext}
We see from Eq. (\ref{eq:WideProbCurrMod}) that one needs to add 
another term $\vct{\mathcal W}_{\mbox{diff}}$ to the probability
current $\vct{\mathcal W}$ when dealing with a SDE. This term does
not exist when dealing with a deterministic equation (Eq.
(\ref{eq:WTotal})), although, the gradient of the forcing entering in the Navier-Stokes equations for $\textbf{A}$ (Eq. (\ref{eq:NS})) has been neglected. The total probability current $\vct{\mathcal W}$
(Eq. (\ref{eq:WideProbCurrMod})) was displayed in Fig.
\ref{fig:RQVec}(h). We would like now to display separately the
probability current coming from the drift terms $\vct{\mathcal
W}_{\mbox{drift}}$ and the one generated by the diffusion
coefficients $\vct{\mathcal W}_{\mbox{diff}}$. We represent in
Fig. \ref{fig:DriDifMod} the vector and streamline plots of the
probability current for the model associated (a) to the drift
coefficients $\vct{\mathcal W}_{\mbox{drift}}$ and (b) to the
diffusion coefficients $\vct{\mathcal W}_{\mbox{diff}}$ (Eq.
(\ref{eq:WideProbCurrMod})). We recall that the total probability
current is displayed in Fig. \ref{fig:RQVec}(h). We see first that
$\vct{\mathcal W}_{\mbox{drift}}$ and $\vct{\mathcal W}_{\mbox{diff}}$ are of the same order of magnitude. The current  $\vct{\mathcal W}_{\mbox{drift}}$ associated to the deterministic part of the stochastic evolution (Eq. (\ref{eq:SysRQSDE})) goes towards the origin in a rotating motion: the dynamics is decaying. To compensate for this decay, the current $\vct{\mathcal W}_{\mbox{diff}}$ associated to the stochastic forcing part of the evolution  (Eq. (\ref{eq:SysRQSDE})) points outwards away from the origin. The sum of these two, the total current displayed in Fig.\ref{fig:RQVec}(h), has a circular motion around the origin, consistent with a stationary process (i.e. $\partial \mathcal P (Q,R)/\partial t =-\vct{\nabla}.\vct{\mathcal W}\approx 0$).


\end{document}